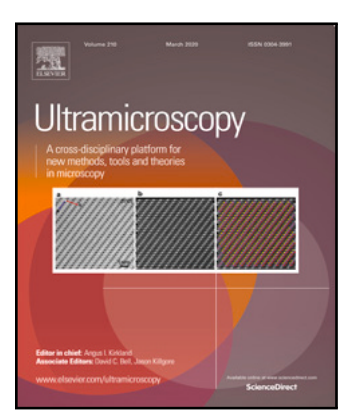

# Orientation-adaptive virtual imaging of defects using EBSD

Nicolò M. della Ventura [a,*], James D. Lamb [a,1], William C. Lenthe [b], McLean P. Echlin [a], Julia T. Pürstl [a], Emily S. Trageser [c], Alejandro M. Quevedo [c], Matthew R. Begley [a], Tresa M. Pollock [a], Daniel S. Gianola [a,*], Marc De Graef [d,*]

[a] Materials Department, University of California, Santa Barbara, Santa Barbara, CA 93106, USA
[b] Gatan + EDAX Inc., Pleasanton, CA 94588, USA
[c] Electrical and Computer Engineering Department, University of California, Santa Barbara, Santa Barbara, CA 93106, USA
[d] Department of Materials Science and Engineering, Carnegie Mellon University, 5000 Forbes Avenue, Pittsburgh PA 15213, USA

ARTICLE INFO

ABSTRACT

Electron backscatter diffraction (EBSD) is a foundational technique for characterizing crystallographic orientation, phase distribution, and lattice strain. Embedded within EBSD patterns lies latent information on dislocation structures, subtly encoded due to their deviation from perfect crystallinity — a feature often underutilized. Here, a novel framework termed orientation-adaptive virtual apertures (OAVA) is introduced. OAVAs enable the generation of virtual images tied to specific diffraction conditions, allowing the direct visualization of individual dislocations from a single EBSD map. By dynamically aligning virtual apertures in reciprocal space with the local crystallographic orientation, the method enhances contrast from defect-related strain fields, mirroring the principles of diffraction-contrast imaging in TEM, but without sample tilting. The approach capitalizes on the extensive diffraction space captured in a single high-quality EBSD scan, with its effectiveness enhanced by modern direct electron detectors that offer high-sensitivity at low accelerating voltages, reducing interaction volume and improving spatial resolution. We demonstrate that using OAVAs, identical imaging conditions can be applied across a polycrystalline field-of-view, enabling uniform contrast in differently oriented grains. Furthermore, in single-crystal GaN, threading dislocations are consistently resolved. Algorithms for the automated detection of dislocation-induced contrast are presented, advancing defect characterization. By using OAVAs across a wide range of diffraction conditions in GaN, the visibility/invisibility of defects, owing to the anisotropy of the elastic strain field, is assessed and linked to candidate Burgers vectors. Altogether, OAVA offers a new and high-throughput pathway for orientation-specific defect characterization with the potential for automated, large-area defect analysis in single and polycrystalline materials.

## 1. Introduction

Over the past two decades, the transmission electron microscopy (TEM) community has developed a series of quantitative observation modalities that are referred to under the umbrella name “4D-STEM” (4D scanning transmission electron microscopy, see [1] for a recent review). These approaches record a 2-D map of intensities in reciprocal space $(q_x, q_y)$, i.e., a diffraction pattern, as a function of the (2-D) beam position $(x, y)$ in real space, hence the “4D” label. Modern pixelated electron detector systems, such as active pixel sensors [2] and hybrid pixel array detectors [3], are ideally suited for these measurements due to their single electron sensitivity, allowing for high mapping speeds with relatively low dose and with high dynamic range, simultaneously recording both the direct and diffracted beams. The ability to store *all* diffraction patterns[2] enables a wide variety of subsequent post-processing techniques, because virtual (digital) apertures and various analysis approaches can be applied to the data interactively. This leads to a multitude of quantitative modalities, including virtual dark field imaging, lattice strain mapping, differential phase contrast, and ptychrography [1,5–8].

On the scanning electron microscopy (SEM) side, three main diffraction techniques are available: electron backscatter diffraction (EBSD), transmission Kikuchi diffraction (TKD) and electron channeling pattern

* Corresponding authors.
*E-mail addresses:* n_dellaventura@ucsb.edu (N.M. della Ventura), gianola@ucsb.edu (D.S. Gianola), degraef@cmu.edu (M. De Graef).
[1] This author contributed equally to this work.
[2] Storage of all the detector data can be considerable in size and pose significant data transfer and handling challenges [4], and has only recently become possible with the advent of relatively inexpensive high-speed solid-state devices.

(ECP) acquisition. In each case, a 2D diffraction pattern is acquired as a function of position $(x, y)$ on the sample, so that these techniques are effectively also examples of 4D approaches (4D-EBSD, 4D-TKD, 4D-ECP), even though they have traditionally not been labeled that way. Nevertheless, the user can store all diffraction patterns so that the 4D dataset can be interrogated using a variety of virtual apertures that target selective diffraction to regions of the detector. Specifically, previous studies [9–13] have introduced a virtual dark-field (DF) imaging technique based on EBSD patterns (EBSPs), sometimes referred to as EBSD-DF, or synthetic or virtual forward or backscatter detectors. This method constructs virtual images by selecting specific locations within the EBSP. To date, static virtual apertures have been employed, where the virtual detector remains fixed with respect to the detector reference frame (or pixel position within). While these static virtual apertures offer flexible and powerful means of contrast generation, they remain fundamentally limited in their ability to isolate diffraction conditions specific to individual {hkl} planes (or Kikuchi bands). In other words, as the virtual aperture is fixed in these implementations, the same region of the detector integrates different diffraction features depending on the local crystal orientation at each point. This leads to a blending of multiple diffraction contributions, thereby obscuring the crystallographic specificity of the generated contrast. As a result, the physical origins of contrast — particularly those arising from crystallographic anisotropy in dislocation or defect visibility — remain entangled and challenging to interpret.

Importantly, successful application of such an EBSD-DF approach hinges on prolonged acquisition times for high-resolution EBSPs, the use of accurate indexing routines, as well as the meticulous determination of the pattern center — a critical requirement for the precise, quantitative interpretation of strain-induced contrast. Recently, pattern reindexing techniques (spherical indexing [14] and dictionary indexing [15,16]) proved that band locations alone (Hough indexing) lead to lower orientation indexing accuracy than using the entire pattern. Moreover, pseudosymmetric indexing confusion is less problematic when EBSD patterns contain higher order band information and is leveraged during indexing [17,18]. Unfortunately, much of the higher order diffraction information in EBSD is obfuscated due to the reduced sensitivity of phosphor based detectors (indirect electron detection [19, 20]) unless prohibitively long exposure times are used.

Silicon-based direct electron detection has been critical to the development of quantitative transmission electron microscopy modalities, particularly 4D-STEM, because of fast patterns acquisition rates (low exposure times) enabled by the enhanced detective quantum efficiency. Similarly high sensitivities have been shown in low-kV adapted detectors for SEM and applied to EBSD data collection [18–23]. Such hardware and software developments point to the tantalizing concept of using virtual apertures that target selective diffraction conditions within the EBSP, thereby isolating any crystal anisotropy that influences the generation of contrast in virtual DF images with high signal-to-noise and high spatial resolution.

For example, the strain fields from lattice defects are encoded in EBSPs in an anisotropic way [24]. Indeed, this forms the basis for dislocation visibility and invisibility in two-beam DF-S/TEM images that are used not only to image a dislocation, but to also determine its Burgers vector when the sample is tilted to a sufficient number of distinct diffraction conditions. Executing this in the context of virtual defect imaging using EBSD requires adaptive placement of the virtual apertures to anchor the specific diffraction conditions in the face of potential changes in crystal orientation. In fact, in the well known ECCI technique (electron channeling contrast imaging) the user sets up a diffraction condition based on either an EBSD pattern [25] or an ECP [26]. In the former technique, the sample is tilted while observing the EBSD pattern until a selected band falls on a diode detector; as the beam is then rastered across the region of interest, an ECCI defect image is acquired for the selected diffraction/channeling condition. In the latter approach, the sample is tilted to bring a particular Kikuchi band onto the optical axis, and then the region of interest is rastered using a beam along that axis. In both cases electrons channel deep into the sample and few backscattered electrons are detected, except for locations where the diffraction/channeling condition is perturbed, i.e., near defects.

In this research effort, dislocations are revealed directly from individual EBSD maps by exploiting orientation-specific diffraction conditions, implemented through a novel approach termed orientation-adaptive virtual aperture (OAVA). This method dynamically positions a virtual aperture in reciprocal space relative to the locally indexed crystallographic orientation, enabling contrast enhancement of the strain field generated by defects. Compared to static aperture placement, OAVA paired with the use of a modern direct electron detector yields markedly improved defect visibility, enabling detailed visualization of individual dislocations across single and polycrystalline samples. Algorithms for the automated identification of dislocation contrast - and therefore the conditions for defect visibility or invisibility criteria - are also presented, providing a path towards large field-of-view automated dislocation analysis.

## 2. Materials and methods

Materials used in the current study comprise (i) poly-crystalline nickel (Ni), and (ii) wurtzite gallium nitride (GaN). For the poly-crystalline Ni sample (Fig. 1a–c), an open source dataset [27] that accompanies a dictionary indexing tutorial paper [16] was used in this work. This dataset consists of three Ni EBSD datasets of 186 × 151 patterns acquired at 20 kV for a sample tilt of 75.7°; the difference between the datasets is the gain setting of the detector, with one dataset, Ni1, consisting of patterns with good signal-to-noise (S/N) ratio, Ni4 with very poor S/N, and the third, Ni6, with intermediate S/N. We will use the dataset labeled Ni1 in this work. The EBSD patterns have dimensions of 60 × 60 pixels (i.e., 8× binning on a 480 × 480 Hikari camera). Further information of the detection parameters used can be found in [16]. Along with the patterns, the open source data repository also provides the end-result of a dictionary indexing run followed by orientation refinement.

The wurtzite GaN sample was prepared via atmospheric pressure metal organic chemical vapor deposition (MOCVD) on 2 inch c-plane (0001)-oriented flat sapphire substrate (FSS) from Cryscore Optoelectronic Limited. The MOCVD growth was performed at the University of California, Santa Barbara (UCSB) on a two-flow reactor using trimethylgalium (TMG) and ammonia ($NH_3$) precursors for Ga and N, respectively, and consisted of a 40 nm low temperature (LT) unintentionally doped (UID) GaN nucleation layer grown at 560 °C, a 2.7 μm high temperature (HT) UID GaN coalescence layer grown at 1220 °C with a $NH_3$ flow of 3 slm and a TMG flow of 15 sccm, targeting a threading dislocation density on the order of $1 \times 10^8\ \text{cm}^{-2}$ [28], and a 2.7 μm HT n-type GaN layer grown at 1220 °C (doping density for Si is in the $1 \times 10^{18}\ \text{cm}^{-3}$ range). This particular GaN sample was chosen for its high threading dislocation density (TDD), as shown in Section S.2 of the Supplementary Materials. The measured TDD on the GaN wafer was roughly $3 \times 10^8\ \text{cm}^{-2}$ (measured using electron channeling contrast imaging — see Figure S.2 in the Supplementary Materials) and is similar to the work described in [29].

For wurtzite GaN, EBSD patterns were collected with the DE-SEMCam manufactured by Direct Electron LP (San Diego, CA USA) using a custom monolithic active pixel sensors (MAPS, full-frame resolution 2048 × 2048, effective pixel size of 13 μm, maximum readout speed 281 fps) [19]. The DE-SEMCam is installed on a Thermo Fisher Scientific Apreo-S scanning electron microscope (SEM) operated at accelerating voltages of 8 and 10 kV. All datasets were collected with raw patterns being saved and exported into a .up2 binary 16-bit format, such that they could be reindexed during post-processing and virtual aperture images could be formed. The complete raw diffraction pattern datasets were acquired at a resolution of 2048 × 2048 with 32-bit images. However, due to the substantial data storage demands associated with the full-resolution datasets (amounting to approximately 16 MB per pattern) and to facilitate data portability, the images were down-sampled to a resolution of 256 × 256 with 16-bit depth during export. While certain applications may benefit from reconstructing virtual images from the full-resolution 2048 × 2048 patterns in future studies, for the materials examined in this work, the down-sampled datasets provided sufficient fidelity for analysis. Further information of the detection parameters used are listed in Table 1. The EBSPs were dark- and flat field-background subtracted. Dark reference backgrounds were collected with the detector in position in the chamber with the electron beam blanked and all photon sources inactive. For flat field-background, dynamic background correction (division) was used for single-crystal GaN.

**Table 1**
Detection parameters for the GaN EBSD scan.

| Material | kV | nA | Fps | Scan size (pts) | Step size (nm) | Scanned area ($\mu m^2$) | Camera tilt |
|---|---|---|---|---|---|---|---|
| GaN | 8 | 3.2 | 25 | 175 × 175 | 20 | 3.5 × 3.5 | 8.5° |
| **Material** | **WD (mm)** | **Pattern size** | **Camera pxl (processed)** | **L (mm)** | **Solid angle** | **PC (x*, y*, z*)** | |
| GaN | 16 | 2048 × 2048 | 256 × 256 | 20.929 | ~65° | 0.5029, 0.3541, 0.7861 | |

For all the EBSD datasets in this work, indexing was performed with EMsoft v6 dictionary indexing [15,16] with the pattern center (PC) determined using the Efit software in EMsoft. These results are used for the orientation-adaptive virtual aperture method described later. An interactive algorithm for selecting static or orientation-adaptive virtual aperture locations and for generating virtual images from EBSD data was developed in C++. Alternatively, a dedicated implementation has been integrated into the EMsoft library through a specialized function called EM4DEBSD [30].

For TEM analysis, an electron-transparent foil was extracted from a random region of the wurtzite GaN sample using a FEI Helios Dualbeam Nanolab 600 focused ion beam (FIB), adhering to established preparation protocols. TEM imaging was conducted at 200 kV on a Thermo Scientific Talos F200X TEM.

## 3. Results and discussion

In the following sections, measurements from two EBSD datasets serve as the basis for demonstrating the orientation-adaptive virtual aperture approach for defect imaging in both polycrystalline Ni and single-crystalline GaN samples (Fig. 1). Fig. 1a and c display the [001] inverse pole figure (IPF) maps for each dataset, Fig. 1b and e present the corresponding normalized dot product (NDP) maps between the experimental and simulated EBSPs, obtained through dictionary indexing, while Fig. 1d reports the local orientation spread (LOS) map for the GaN dataset. Notably, Fig. 1d and e reveal the presence of surface-penetrating defects (threading dislocations) within the GaN sample, while in Fig. 1b grain-scale and subgrain-scale diffraction contrast is present.

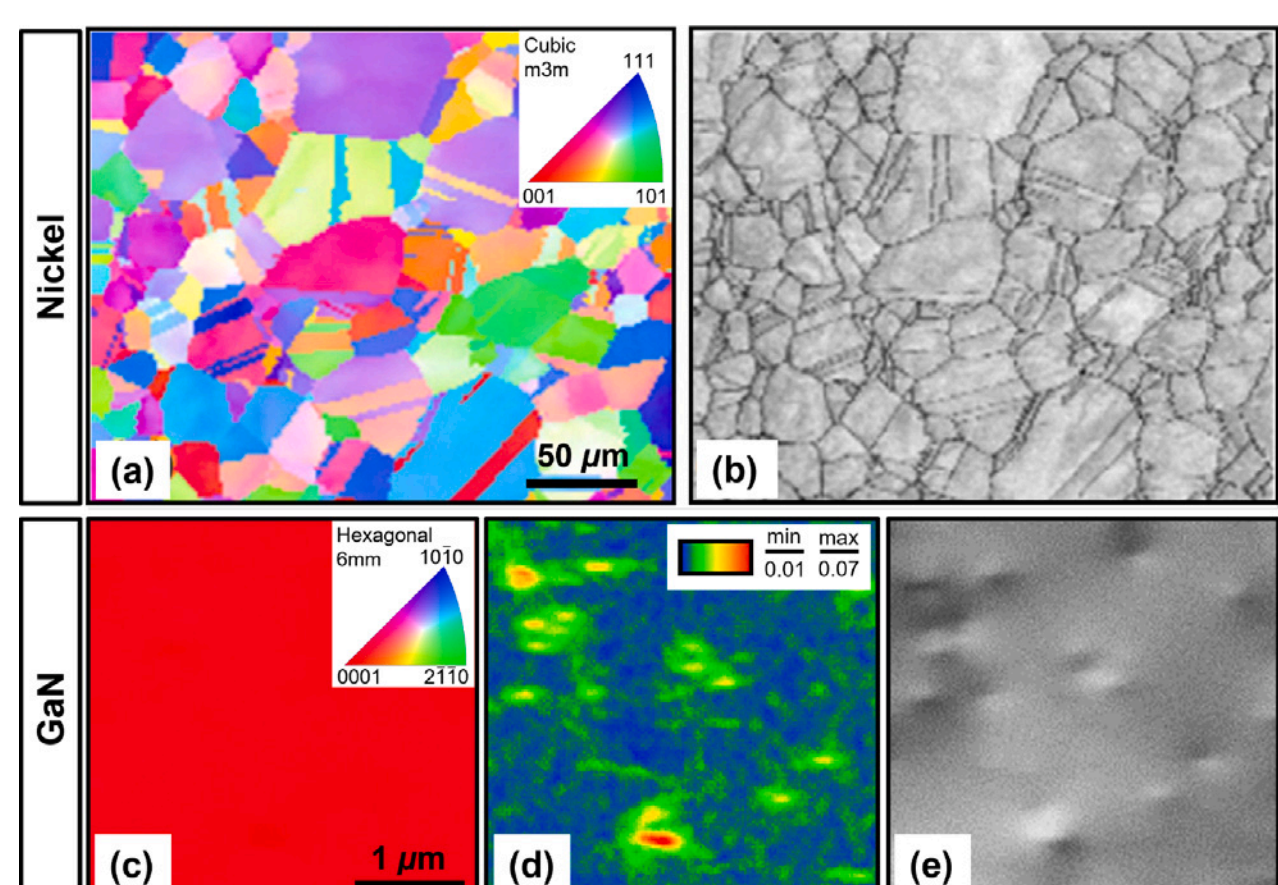

**Fig. 1.** Overview of the polycrystalline Ni and single crystal GaN samples used in this work. (a, (c) [001] inverse pole figure (IPF) maps; (b, e) normalized dot product (NDP) maps; (d) local orientation spread (LOS) map for the GaN dataset, computed based on the fourth-nearest neighbor. The Ni dataset comprises 186 × 151 patterns, each with a resolution of 60 × 60 pixels. The GaN dataset comprises 175 × 175 patterns, each with a resolution of 256 × 256 pixels.

### 3.1. Defining virtual apertures: approach and computational framework

In a conventional EBSD workflow (Fig. 2), a 2D map is generated by scanning an electron beam across a tilted crystalline sample, where backscattered electrons form Kikuchi patterns on the detector. Storing all EBSD patterns enables post-processing to extract various image representations, such as IPF and NDP maps (as reported in Fig. 1 for Ni and GaN). Yet, since an EBSD dataset includes a full-pattern at each sampling location, alternative image generation modalities can be employed, such as using a virtual aperture at any position within the pattern. A virtual detector image is generated by integrating the intensity over the virtual aperture area for all patterns in the dataset. Any virtual aperture shape can be used since it is digitally constructed, such as square or circular, with uniform or Gaussian weight profiles across the aperture. In this work, a Hann window profile within a square virtual aperture is used since the weights go to zero at the edge of the square. The mathematical expression for the 2-D Hann window is given by:

$$H(x,y) = (\alpha(1-\cos(2\pi x)) + \cos(2\pi x)) \times (\alpha(1-\cos(2\pi y)) + \cos(2\pi y)), \quad (1)$$

where $x$ and $y$ lie in the interval $[-\frac{1}{2}, \frac{1}{2}]$; the function goes to zero at the edge of the interval when $\alpha = \frac{1}{2}$. The coordinate values can be scaled to obtain a virtual aperture of arbitrary size.

Regardless of its shape, a virtual aperture can be efficiently applied to an EBSD dataset through convolution using fast Fourier transforms (FFT), enabling rapid processing of EBSPs across the full dataset. The convolution operation can be parallelized, ensuring computational efficiency. Importantly, this operation generates image intensities corresponding to the same physical sampling points as in the original scan.[3] By performing bilinear interpolation on the convolved diffraction patterns, a virtual image of the scanned material can be reconstructed for any chosen virtual aperture placement within the EBSPs. This enables flexible image generation from an arbitrary virtual aperture position. When single-pixel-sized virtual apertures are employed, the application of windowing functions (e.g., Hann window) and convolution operations becomes unnecessary, as the intensity at each aperture position can be directly extracted from the data without additional spatial weighting or integration. This simplification reduces computational overhead while ensuring a direct correspondence between the measured and reconstructed intensities. To ensure that the virtual image accurately represents diffraction contrast rather than background intensity variations in the EBSPs, background subtraction/division is performed before applying convolution.[4]

[3] When the output is cast to the same data type as the original dataset, the resulting dataset maintains the same dimensions as the original one.

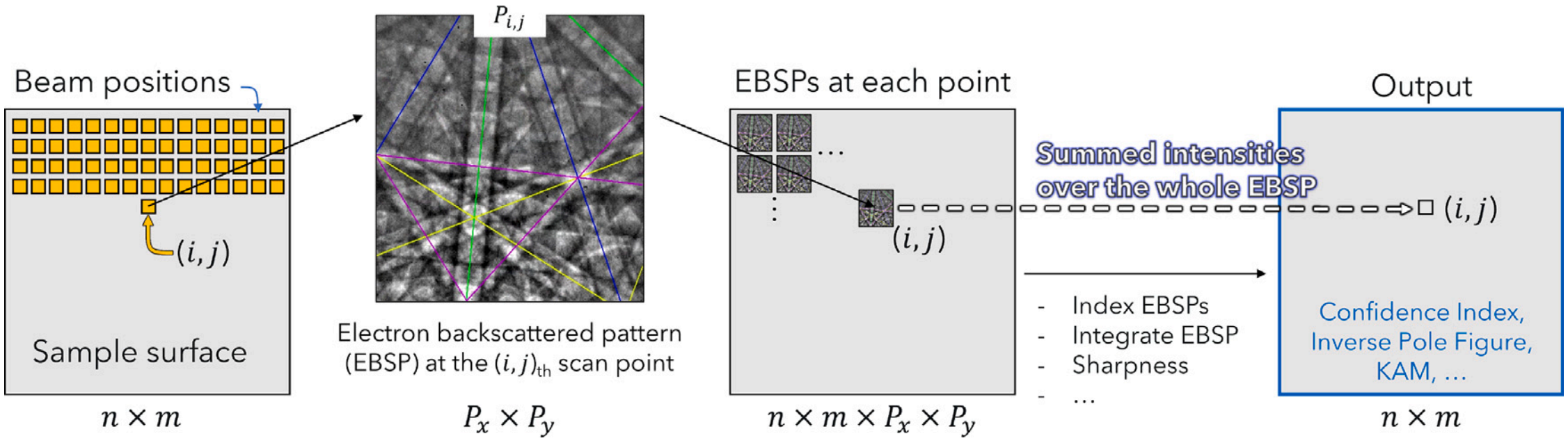

**Fig. 2.** A conventional EBSD workflow where a 2D map of $n \times m$ points is collected. At each sampled point in the scan there exists a 2D EBSD pattern of $P_x \times P_y$ pixels. By storing all patterns during mapping, various image representations can be generated through post-processing, including image quality maps, confidence index maps, pattern sharpness maps, and the indexing or re-indexing of crystallographic orientations.

Expanding this approach, the original dataset can be interrogated in a number of ways, including: (1) a *periodic array of static virtual apertures* in which each virtual aperture is fixed to a $(p_x, p_y)$ position of each EBSP for the entire EBSD dataset. Each virtual aperture generates a distinct virtual image (Fig. 3a). Using only a single virtual aperture out of the entire array is equivalent to forming a virtual image with a stationary detector at some defined location in the diffraction pattern. In this case, the position of the virtual aperture remains invariant with respect to changes in crystal orientation across the scanned area. As this approach does not compensate for orientation-induced shifts in the diffraction pattern, it potentially leads to inconsistencies in the virtual image. Alternatively, (2) a set of *orientation-adaptive virtual apertures* in which the virtual aperture position within each EBSP is adaptively (dynamically) computed based on the sample symmetry and the orientation derived from each individual EBSP (Fig. 3b). The OAVA-based imaging approach introduced here, allows for the dynamic targeting of specific diffraction imaging conditions. By accounting for orientation variations and potential shifts in the pattern center, this method ensures that the same diffraction condition is consistently achieved, thereby improving the accuracy and reliability of the virtual image formation and its application to defect analysis. A comprehensive description of these two methods is presented in the following sections.

### 3.1.1. Periodic array of static virtual apertures

We first use static virtual apertures as the basis for comparison with our OAVA approach. Starting with a convolved pattern file, a periodic 2D array of convolved pattern intensities can be extracted based on a chosen virtual aperture step size and mapped to the appropriate locations within an array of virtual aperture images. Given an EBSD map consisting of $n{\times}m$ points, each point is associated with a diffraction pattern of resolution $P_x \times P_y$ pixels. A virtual aperture array is applied to each pattern with a defined step size of $v$ pixels. The first virtual aperture is positioned at coordinates $(v/2, v/2)$. Under these conditions, the number of virtual aperture positions along each pattern dimension is determined by the integer divisions $P_x/v$ and $P_y/v$, respectively. This results in an overall periodic array image of size $R \times Q$ pixels, with $R = (P_x/v) \times n$ and $Q = (P_y/v) \times m$. That is, $R$ and $Q$ are dictated by the spatial arrangement of the virtual apertures and the sampling points. Given that this approach has been previously introduced in the literature for a single virtual aperture location choice [9–12], the results of this method are presented in Section S.1 of the Supplementary Materials for completeness.

[4] Bit-depth reduction was performed after background correction to preserve the dynamic range of the GaN patterns.

### 3.1.2. Orientation-adaptive virtual apertures (OAVA)

We next describe the novel orientation-adaptive virtual aperture (OAVAs) approach, which dynamically adjusts the virtual aperture position so as to be linked to a specific diffraction condition (and Kikuchi band) across all patterns in the EBSD map. Specifically, the aperture placement is determined adaptively based on the crystal symmetry and orientation extracted from the indexing of each individual mapping location. This methodology enables a more targeted interrogation of the original dataset, leveraging the intrinsic crystallographic information to generate virtual images tied to a specific set of crystal planes $hkl$.

*Computational framework*

The computational framework for the OAVA approach begins with the convolved pattern file and a simulated Kikuchi sphere for the material being studied. This can take the form of a single intensity distribution on the surface of the sphere, calculated for the exact accelerating voltage of the microscope [31], or a series of concentric Kikuchi spheres calculated for a discretized energy range using a combined Monte Carlo and dynamical scattering approach [32]. For numerical convenience, the intensity distribution is represented as a stereographic projection, but this is not essential for the proposed algorithm. The intensity distribution is represented by the symbol $\mathcal{M}(\theta, \varphi)$ where the arguments denote the usual spherical angles; $\theta \in [0, \pi]$ starting from the North pole of the projection, which coincides with the intersection of the reciprocal $\mathbf{c}^*$ axis with the projection sphere, and $\varphi \in [0, 2\pi]$ where the value of 0 corresponds to the intersection of the crystallographic $\mathbf{a}$ axis with the sphere in the equatorial plane.

An example stereographic projection for Ni at 20 kV accelerating voltage is shown in Fig. 4(b) along with the corresponding unit radius Kikuchi sphere in (a). As a representative case, a near two-beam diffraction condition is selected as the crystal direction of interest. This condition corresponds to darker regions within the diffraction pattern, positioned near Kikuchi bands yet away from the major zone axes. An instance of such a near two-beam diffraction condition - characterized by a pronounced contrast variation across a Kikuchi band edge - can be observed in proximity to the [101] zone axis, as indicated by the yellow arrow in the inset of Fig. 4b. If we represent the stereographic coordinates of the selected point by $(X_s, Y_s)$ (in units of pixels with respect to a 2D cartesian reference frame located at the center of the projection), then this corresponds to a unit vector $\hat{\mathbf{n}}$ on the Kikuchi sphere with components:

$$\hat{\mathbf{n}} = (\sin\theta\cos\varphi, \sin\theta\sin\varphi, \cos\theta) \tag{2}$$

with

$$\theta = \arccos\left(\frac{1 - \bar{X}_s^2 - \bar{Y}_s^2}{1 + \bar{X}_s^2 + \bar{Y}_s^2}\right), \varphi = \arctan\left(\frac{\bar{Y}_s}{\bar{X}_s}\right),$$

and $(\bar{X}_s, \bar{Y}_s) = (X_s, Y_s)/N$ with $N$ the number of pixels corresponding to the projection circle radius.

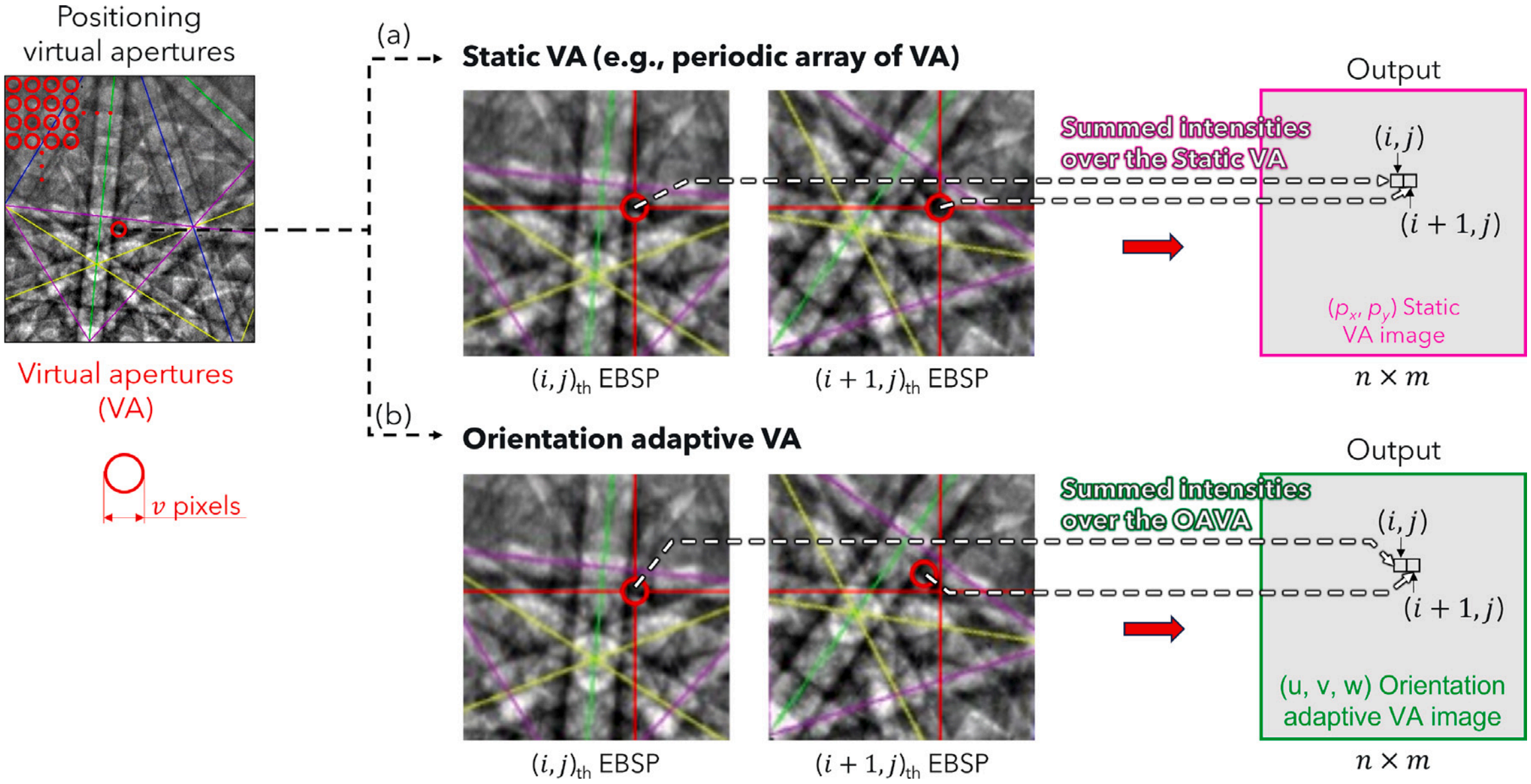

**Fig. 3.** Schematic illustration of the virtual imaging process using static and orientation-adaptive virtual apertures (OAVA). (a) The static virtual aperture approach, wherein the virtual aperture is fixed at a position $(p_x, p_y)$ within the EBSD pattern and remains unchanged across all scanned points, irrespective of variations in the crystallographic orientation of the sample. (b) The orientation-adaptive virtual aperture approach, wherein the virtual aperture is dynamically adjusted to follow a specific diffraction condition (or crystal direction) across the scanned area.

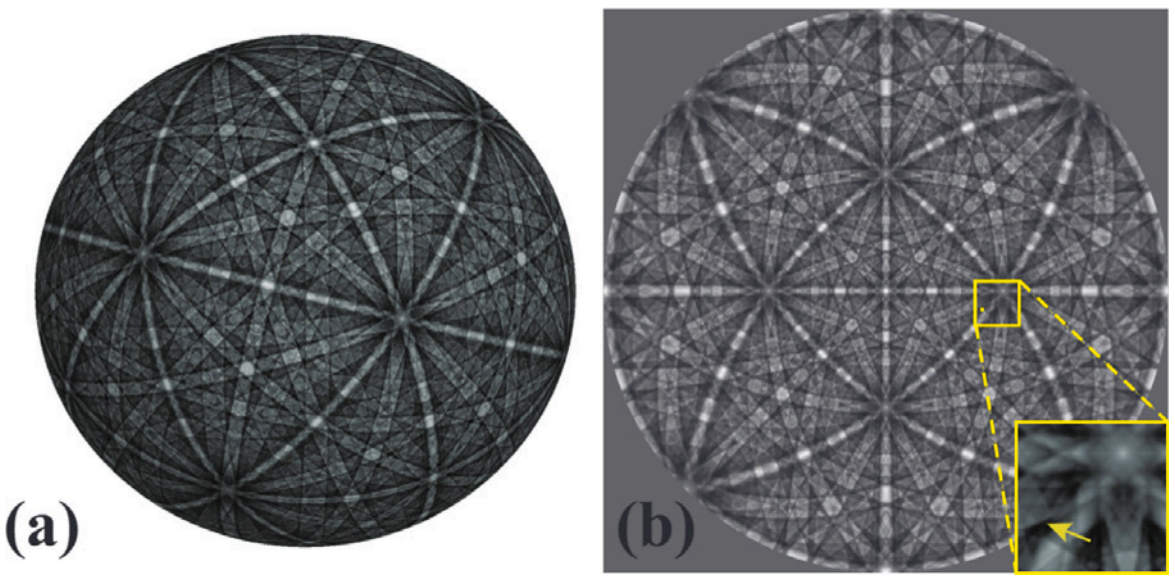

**Fig. 4.** (a) Kikuchi sphere for Ni at 20 kV accelerating voltage simulated with EMsoft; (b) corresponding stereographic projection of the northern hemisphere. The inset in the lower right corner shows an example of a rapid contrast change close to a zone axis which generally corresponds to a local near two-beam diffraction condition.

If the lattice orientation at a particular sampling point is known and represented by a (passive) unit quaternion $q$, then the unit vector $\hat{\mathbf{n}}$ can be rotated into the sample reference frame by means of the standard quaternion operation $\hat{\mathbf{n}}_s = V[q[0, \hat{\mathbf{n}}]q^*]$, where the subscript $s$ stands for the sample reference frame, the asterisk denotes quaternion conjugation, and the versor $[0, \hat{\mathbf{n}}]$ is a quaternion with zero scalar part. The operator $V$ extracts the vector part of its quaternion argument. The resulting vector $\hat{\mathbf{n}}_s$ represents a line that intersects the detector plane in a point $(x_d, y_d, z_d)$ (expressed in the sample reference frame) given by:

$$\begin{pmatrix} x_d \\ y_d \\ z_d \end{pmatrix} = \mathcal{T} \begin{pmatrix} n_{s,x} \\ n_{s,y} \\ n_{s,z} \end{pmatrix} \tag{3}$$

with

$$\mathcal{T} = \begin{pmatrix} x_{pc}\delta \sin\alpha & -L & x_{pc}\delta\cos\alpha \\ y_{pc}\delta\sin\alpha - L\cos\alpha & 0 & L\sin\alpha + y_{pc}\delta\cos\alpha \\ \sin\alpha & 0 & \cos\alpha \end{pmatrix} \tag{4}$$

This relation was obtained by inverting eq. (11) in [32], with $(x_{pc}, y_{pc})$ as the pattern center coordinates expressed in units of pixels with respect to the center of the detector, $\delta$ is the detector pixel size in microns, and $L$ is the normal distance from the illumination point to the detector plane (in microns). The angle between ND (normal direction in the sample's reference frame RD, TD, ND — rolling, transverse, and normal directions, respectively) and the scintillator normal is $\alpha = \pi/2 - \sigma + \theta$, with $\sigma$ the sample tilt angle measured from horizontal and $\theta$ the detector tilt angle measured from vertical. A final transformation of $(x_d, y_d, z_d)$ to the detector pixel coordinates $(x, y)$ with respect to the detector center results in:

$$x = \frac{y_d}{\delta} - \frac{1}{2}(N_x - 1) - x_{pc}; \tag{5}$$

$$y = \frac{x_d}{\delta}\cos\alpha - \frac{L}{\delta}\tan\alpha + \frac{1}{2}(N_y + 1) - y_{pc}. \tag{6}$$

The transformation sequence $(X_S, Y_S) \rightarrow \hat{\mathbf{n}} \rightarrow \hat{\mathbf{n}}_s \rightarrow (x_d, y_d, z_d) \rightarrow (x, y)$ is then used to determine where a crystal direction $\hat{\mathbf{n}}$, defined by the user as the point $(X_S, Y_S)$ on the stereographic EBSD master pattern, will intersect the detector plane for a given grain orientation $q$. For a sample region with an orientation gradient this procedure thus allows for the tracking of the diffraction condition defined by $\hat{\mathbf{n}}$ across the detector as the lattice orientation gradually changes, thereby maintaining a constant diffraction condition, whence the name "orientation-adaptive virtual aperture".

If $\mathcal{G}$ represents the (rotational) point group of the sample and $M$ is the group order, then there are at most $M$ symmetrically equivalent unit vectors $\hat{\mathbf{n}}_i$ $(i \in [1 \ldots M])$. We will assume that a fully indexed EBSD dataset of $n \times m$ patterns is available; $(n, m)$ are the width and height of the inverse pole figure map in units of pixels. For each point $(i, j)$, with $i \in [1 \ldots n]$ and $j \in [1 \ldots m]$, an indexed EBSD pattern $P_{i,j}$ is available with a corresponding orientation quaternion $q_{i,j}$ which expresses the (passive) rotation that takes the sample's reference frame (RD, TD, ND) into the crystallographic cartesian reference frame. For a given EBSD detector geometry one can then determine where each of the $M$ symmetrically equivalent unit vectors $\hat{\mathbf{n}}_i$ intersects the detector plane, using the procedure described in the previous paragraph. If a virtual aperture is placed at one of the locations $\hat{\mathbf{n}}_k$, then the position of this virtual aperture in the EBSD pattern can track any local orientation changes inside a given grain, or even larger orientation changes across a grain boundary. Since there are up to $M$ equivalent positions to place the virtual aperture, one can select the position that is closest to the pattern center, or closest to the center of the detector where the background intensity is often highest. However, depending on the values of $L$ and $\delta$, i.e., on the detector solid angle, it is possible that for a collected pattern, say $P_{i,j}$, no virtual apertures for a chosen $\hat{\mathbf{n}}$ falls inside the physical detector area; in such a case, the intensity of the (i, (j) $_{th}$ scan point in the virtual image generated from the virtual aperture tied to the crystal direction $\hat{\mathbf{n}}$ cannot be determined and would result equal to zero.

Another possible reason for not finding any virtual aperture location on a given EBSP is the presence of pseudo-symmetry caused by the absence of an inversion center in the crystal structure. Generally, the absence of an inversion center results in subtle intensity differences between the northern and southern hemispheres of the Kikuchi sphere and pattern indexing may not always produce the correct orientation. Since the orientation-adaptive virtual aperture approach employs the northern hemisphere to identify the diffraction condition, then imposing an inversion center may resolve the issue of the algorithm not locating a valid diffraction condition, even when the crystal structure does not have an inversion center.[5]

*Application of the OAVA approach*

We first show a proof-of-concept demonstration of the OAVA approach for the Ni dataset (Fig. 5). In Fig. 5a, a near two-beam diffraction condition is selected on the stereographic projection of the master pattern (red square) near the [001] zone axis orientation. The symmetrically equivalent locations are highlighted on the master pattern as white boxes. The selected location has pixel coordinates $(9, 33)$ with respect to the center of the stereographic projection which has a size of 500 pixels. In Fig. 5b, the EBSP is shown for an arbitrary scan point selected in the [001] IPF map shown in Fig. 5d; the point is highlighted in white in the upper central grain and has pixel coordinates $(90, 20)$ with respect to the upper left corner of the IPF map.[6] Using the transformations from stereographic coordinates to detector coordinates for the orientation of the selected point, one can determine the location on the detector that corresponds to the diffraction condition selected on the master pattern; the selected sampling point is number $3,624 = (20-1)\times 186+90$ out of 28,086 total patterns and the refined orientation for this point is given by the Bunge Euler angle triplet $(5.974, 0.599, 0.905)$ in radians or quaternion $q = [0.9134, -0.2423, 0.1683, -0.2804]$. The symmetrically equivalent virtual aperture locations for the high cubic rotational point group are highlighted as white squares on the EBSP in Fig. 5b and the point closest to the center of the detector, highlighted in yellow, is selected as the active single-pixel-sized virtual aperture. The detector intensity is then determined by bilinear interpolation from the corresponding pattern within the convolved pattern file and transferred to the virtual aperture image at the location of the selected pixel from Fig. 5d.

Repeating this process for all sampling points in the EBSD map generates the orientation-adaptive virtual dark-field image shown in Fig. 5c, where black regions correspond to grain orientations for which the selected diffraction condition is not present in the EBSP. Note that the intensity is roughly constant in many of the grains and the intensity variations between grains are rather small; in fact, in several grains, the contrast between twin variants nearly completely vanishes. The intensity variations across grains will become smaller for microstructures with larger grains, and, ideally, would vanish for large areas with a uniform orientation or a small orientation gradient. The OAVA approach is therefore particularly well-suited for polycrystalline materials, as it enables orientation-sensitive imaging without the need for specimen tilting.

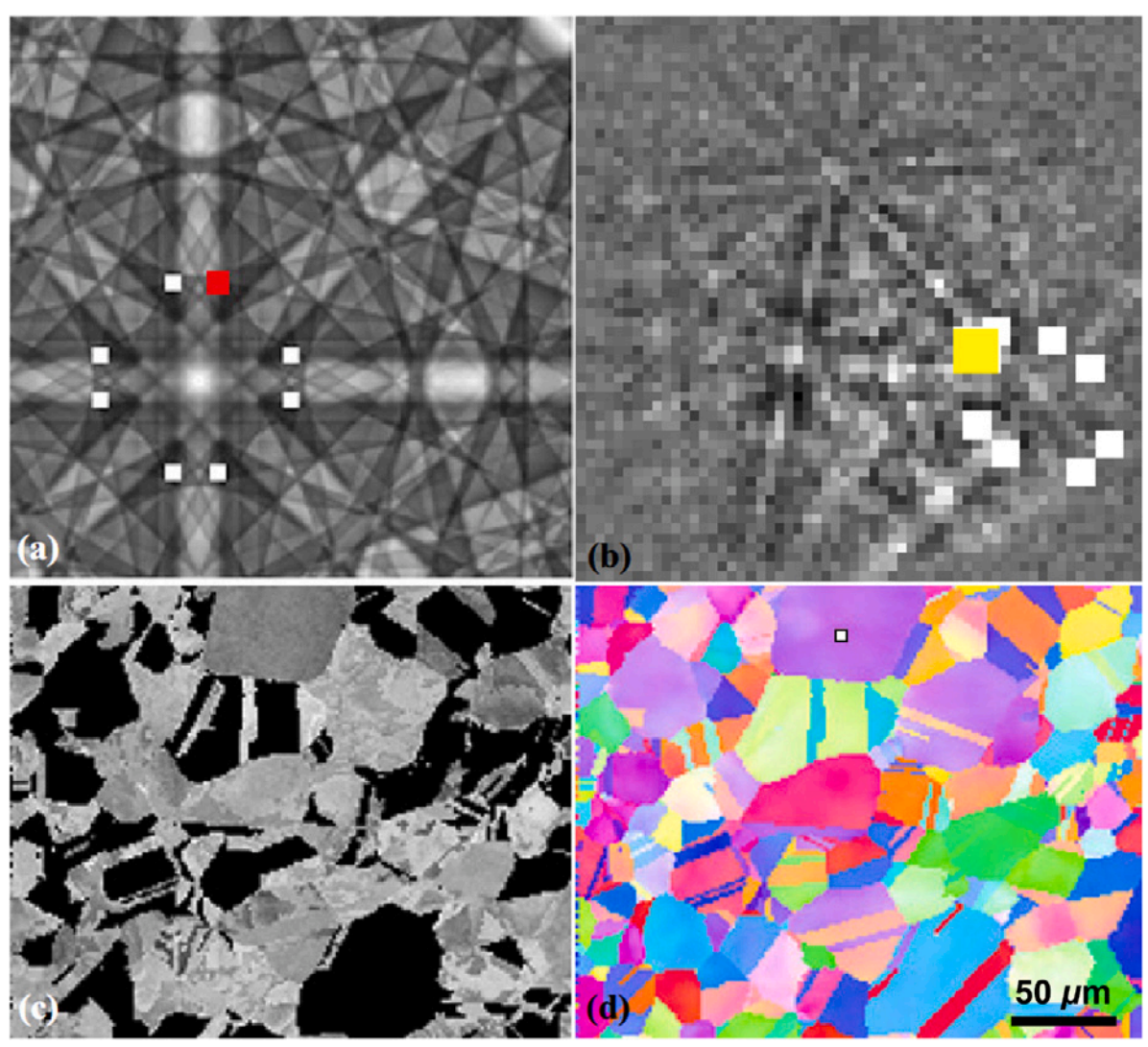

**Fig. 5.** (a) Central portion of the stereographic projection of the Kikuchi sphere for the polycrystalline Ni sample, with a selected diffraction condition indicated in red and its symmetric equivalencies shown in white; (b) the EBSP from the location marked in white near the top-center of the IPF map in (d), with the OAVA diffraction condition now shown in yellow and the symmetric equivalencies in white; (c) the orientation-adaptive virtual aperture image formed by the aperture location in yellow and red. The virtual image is generated using a single-pixel-sized virtual aperture. (d) [001] IPF map.

### 3.2. *Defect imaging with orientation-adaptive virtual apertures*

We next demonstrate how the use of OAVAs applied to dislocations provides a rich set of virtual images linked to specific diffraction conditions. If surface-penetrating defects are present within the imaging area and the interaction volume is an appropriate size [24], then the OAVA approach can be employed for defect imaging. Given that defect imaging necessitates high pattern quality, EBSPs acquired with direct electron detectors dramatically improve the results. We present an analysis of the GaN dataset under varying diffraction conditions through the implementation of OAVAs. To minimize the influence of pattern center variations across the analyzed region, high magnification was employed to confine the imaging area to a sufficiently small field of view. Alternatively, algorithms to dynamically correct the pattern center location may be implemented to enable scalable application of the subsequent analysis.

Fig. 6a,b shows an experimental and simulated pattern for the GaN sample. The simulated pattern is from a dynamical master pattern simulation for GaN at 8 kV visualized using a gnomonic projection in EMsoft [33]. The experimental pattern in Fig. 6a is an EBSP from within the scanned area with an orientation closest to the average orientation of the entire scan. The PC location is indicated with a white plus symbol in the simulated pattern. In Fig. 6(c), the entire EBSD map area is shown using the normalized dot product, which reveals a number of bright/dark contrast regions that are indicative of the strain field surrounding surface penetrating defects. Two different single-pixel-sized orientation-adaptive virtual apertures are applied to this map, using the diffraction conditions indicated with the purple and yellow points in Fig. 6(a), resulting in the images formed in Fig. 6(d,e). The diffraction conditions chosen are at each side of the $(2\bar{1}\bar{1}0)$ band edge, and result in images with visible dislocation contrast. Notably, contrast inversion is observed across the Kikuchi band for all the defects. This behavior is in general expected, as positioning the virtual aperture on opposite sides of a Kikuchi band corresponds to changing the sign of the excitation error, $s_{\mathbf{g}}$ [34]. These virtual images demonstrate the sensitivity of the virtual image contrast to the specific diffraction conditions used, which are targeted via the use of the OAVAs.

[5] Wurtzite GaN used in this study is non-centrosymmetric (i.e., lacks inversion center).

[6] Note that orientation similarity maps can be employed to visualize the degree of similarity between the dictionary indexing results of a given pixel and its nearest neighbors, thereby clearly delineating the grain boundaries. This technique can be integrated into the OAVA approach to help in accurately determine the spatial coordinates of specific grains.

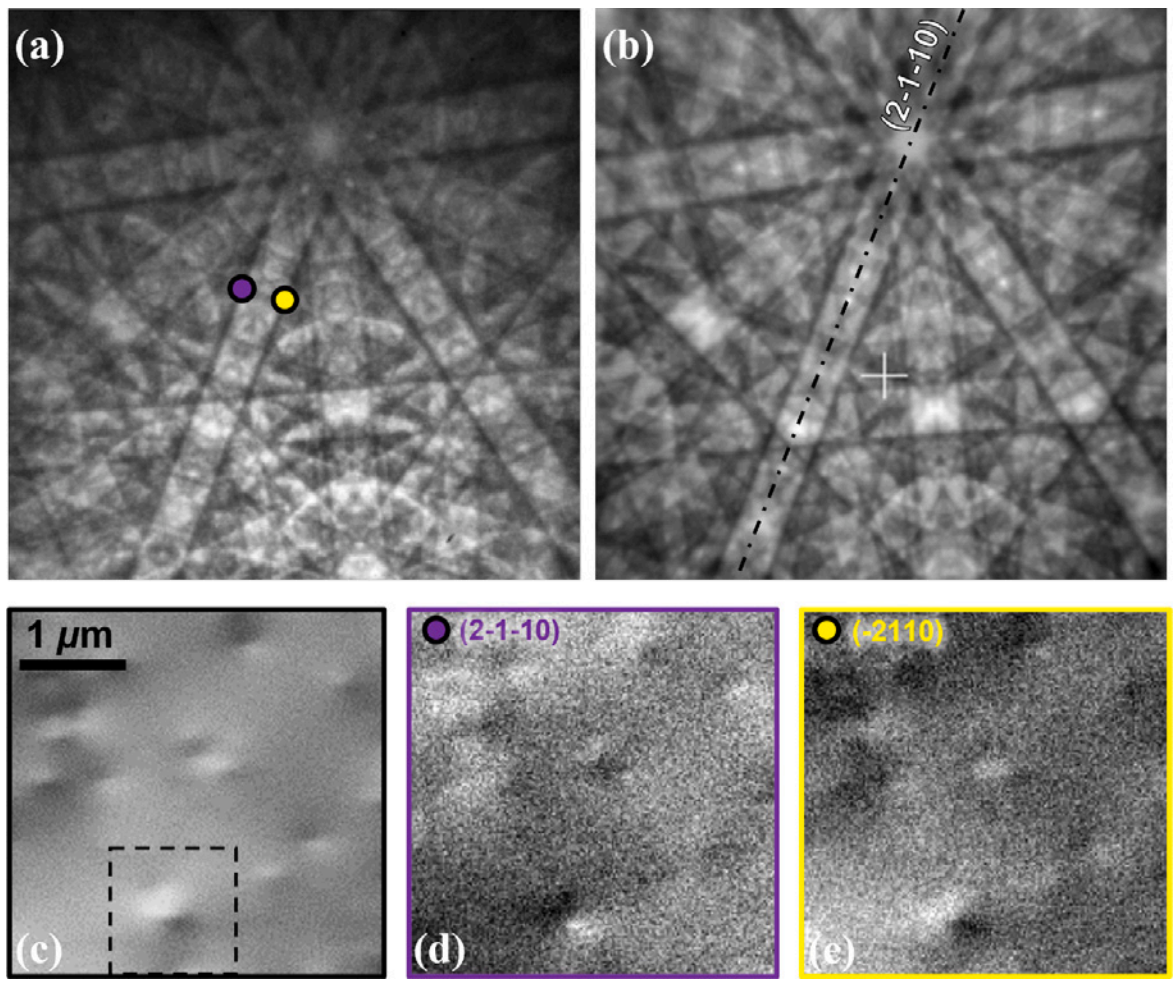

**Fig. 6.** (a) Experimental EBSD pattern from a GaN sample with orientation that is roughly the average of the map, and (b) the corresponding simulated master pattern from EMsoft (pattern center at white plus). (c) Normalized dot product map from Fig. 1e for comparison with the orientation-adaptive virtual aperture images in (d,e) generated using the single-pixel-sized OAVAs located at the diffraction conditions indicated in (a) at the purple and yellow points placed along the $(2\bar{1}\bar{1}0)$ and $(\bar{2}110)$ band. The specific band that was used for image formation is annotated in (d) and (e).

### *3.2.1. Automated defect identification method*

OAVAs allow for forming images with a extensive range of diffraction conditions, all without the need for sample tilting as is required in diffraction-contrast S/TEM and ECCI. To aid in the automated identification and quantification of defect contrast in EBSD-derived virtual aperture images, an algorithm is employed that emulates the structure of strain fields surrounding dislocations. Specifically, a centrosymmetric kernel convolution characterized by two Gaussian distributions centered around each pixel is implemented to reflect the dipolar strain field of a surface penetrating dislocation. This kernel is applied across the entire virtual image generated by the OAVA technique. By sweeping the kernel over the image, the algorithm quantifies both the contrast gradient and a vector indicating the direction of the maximum contrast gradient at every pixel in the image. In the following, this method is applied to an isolated defect shown in the dashed black square in Fig. 6c to test the efficacy and robustness of our method. A kernel with a dimension of 17 pixels is used, as illustrated in the inset in Fig. 7a. A resulting map of the maximum contrast gradient (MCG) can then be generated by calculating each pixel surrounding the dislocation (Fig. 7b). The location of the peak contrast gradient in the MCG map indeed corresponds to the position of the defect (defined as the centroid of the dipolar strain field). Additionally, a maximum contrast gradient direction (MCGD) map (Fig. 7c) shows the angle associated with the pixel exhibiting the most significant contrast gradient in the MCG map. These two metrics capture the centroid of the dislocation and the angle of the vector connecting the bright/dark contrast generated by the defect's strain field.

The defect identification method is validated by generating virtual images for five single-pixel-sized apertures positioned across a band edge (Fig. 8). As the virtual aperture location shifts from outside to inside the band (left to right of the band edge), the defect contrast intensity varies substantially. The image at the green position, with the virtual aperture located just outside the band edge, exhibits the highest defect contrast, while the defect contrast progressively diminishes as the aperture moves further inside the band, becoming nearly imperceptible (invisible) at the red aperture position. These results can be reconciled by considering the excitation error or deviation parameter, $s_\mathbf{g}$ of the location of the virtual aperture, which quantifies the reciprocal distance from the ideal Bragg condition for a given reflection **g** as measured by the distance of the reciprocal lattice point **g** from the Ewald sphere along the sample normal [34]. A positive $s_\mathbf{g}$ corresponds to the reciprocal lattice point being inside the Ewald sphere, while a negative value indicates it is outside, with a zero value representing the exact Bragg condition. When absorption is included, dynamical scattering theory predicts an asymmetry between the $s_\mathbf{g} > 0$ and $s_\mathbf{g} < 0$ conditions, with *anomalous transmission* occurring for the former and *anomalous absorption* for the latter, resulting in the strongest defect diffraction contrast for $s_\mathbf{g} > 0$ [34]. In this condition, the backscattering yield is minimal, leading to the optimal channeling condition, consistent with the virtual image generated using the green OAVA.

### *3.2.2. Defect analysis*

With the defect contrast gradient strongly influenced by the positioning of the virtual aperture on the EBSP, we expand the previous analysis to encompass all virtual images generated using single-pixel-sized virtual apertures positioned at every crystallographic direction accessible on the pattern, with the objective of uncovering systematic dependencies between defect contrast and specific scattering vectors, thereby providing insight into the Burgers vector of the underlying defect.

A 256 × 256 array of OAVA images was generated for a 30 × 30 pixel region surrounding an isolated defect. The full-resolution, uncompressed image is available for download in the supplementary materials ("Mosaic-256 × 256-OAVAs"). This array of OAVA images accounts for all accessible diffraction conditions available using this specific direct EBSD detector and sample-to-detector geometry and distance. It is important to note that virtual apertures located near the detector's edges (i.e., corresponding to the limits of the sampled reciprocal space) can produce undefined contrast in the virtual images. This arises when local orientation changes, for example those surrounding dislocations, displace the virtual aperture beyond the boundaries of the experimentally sampled reciprocal space. In such cases, the resulting virtual images exhibit signal loss, which reflects the absence of a physically meaningful diffraction signal in that region. In Fig. A.1, we present the 254 × 254 array of OAVA images, excluding the edge apertures affected by this effect. Upon contrast normalization, the exclusion of the edge apertures enhances the visualization of the isolated defect in the virtual images.[7]

The centrosymmetric kernel convolution approach is thus applied to each one of the 254 × 254 virtual images to map the defect visibility as a function of the diffraction condition, resulting in an equal number of MCG and MCGD maps. By extracting the peak value from each of these 254 × 254 MCG maps, we compute maps of the peak contrast gradient (PCG) values and peak contrast gradient direction (PCGD) for each pixel of the EBSP for the selected defect (Fig. 9b and c, respectively). A composite map superimposing the PCG and PCGD values for the examined defect reveals several interesting features and a striking sensitivity to the specific diffraction conditions used for virtual imaging (Fig. 9c).

Specifically, the results in Fig. 9b,c reveal that the defect contrast is most pronounced along the band edges, as expected, with the intensity diminishing as one moves away from them. Furthermore, the interaction of intersecting bands introduces variations in contrast intensity when moving along a band, owing to the increased number of crossing diffracted beams, either enhancing or diminishing the defect contrast. Notably, Fig. 9c reveals that a sign inversion in the defect contrast gradient direction occurs when transitioning across any Kikuchi band, whilst no inversion is noted on either side of the zone axes. Interestingly, Fig. 9b,c reveals that for specific Kikuchi bands (i.e., for specific **g** vectors) either no discernible contrast gradient or only a faint modulation is observed along band edges. A clear instance of this occurs for the entire (0$\bar{1}$10) Kikuchi band (indicated in Fig. 9a). The absence of defect contrast along an entire Kikuchi band can be directly interpreted as an indicator of the defect's invisibility condition for a specific **g** vector. Accordingly, Fig. 9b and c illustrate a high-throughput approach for identifying the diffraction conditions under which a defect is either visible or invisible, thereby offering a means to determine the defect's Burgers vector directly from EBSD. Motivated by this result, we next propose an approach for determining the Burgers vector of the selected defect, demonstrating the potential of the OAVA approach for defect analysis.

[7] Note that in the specific case presented in this work, characterized by isolated defects and relatively small orientation gradients, the signal loss near the detector edge affects only a single row of pixels, specifically those corresponding to apertures nearest the detector boundaries. However, in cases involving larger misorientation variations, a broader region adjacent to the detector edge may exhibit signal loss in the resulting virtual images.

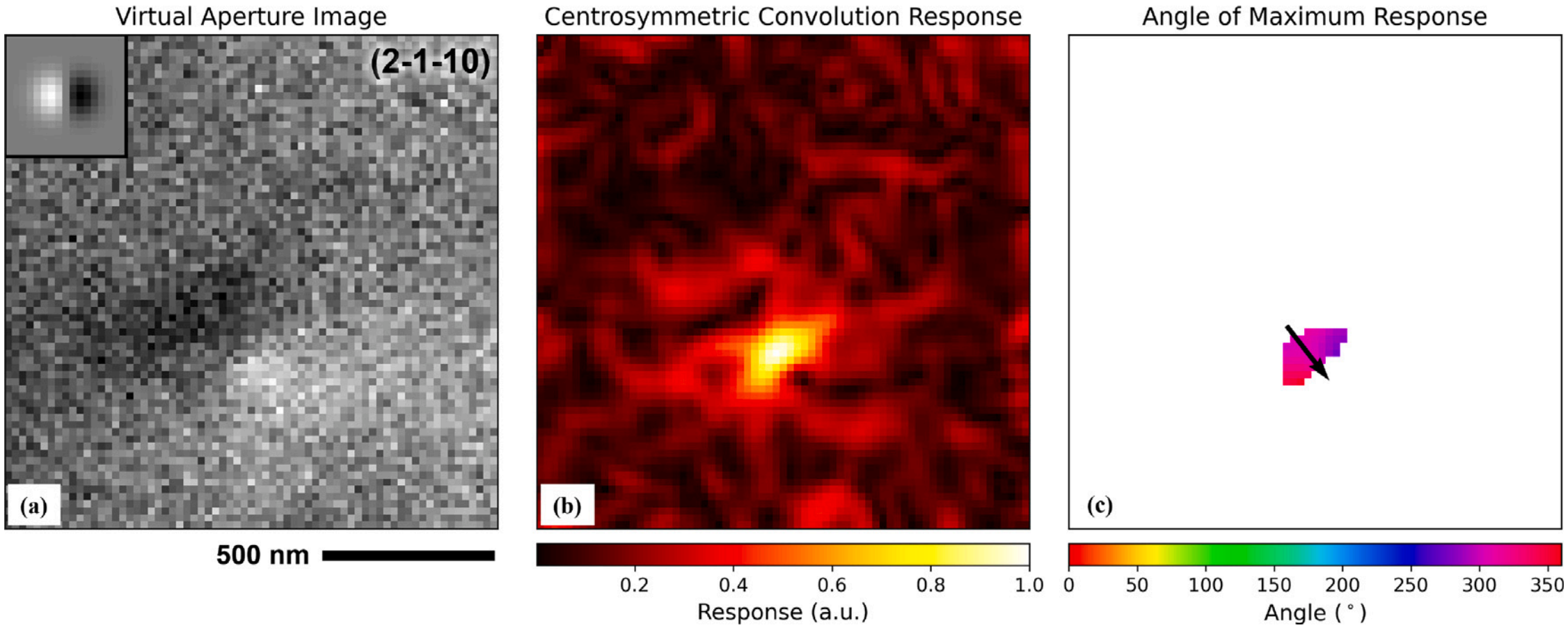

**Fig. 7.** (a) Convolutional kernel used to identify the presence and location of strain fields surrounding a threading dislocation in GaN. The specific band that was used for image formation is annotated in (a). The convolution approach quantifies the magnitude (in arbitrary unit) of the maximum contrast gradient at each pixel, revealing the presence or absence of defect contrast (b), while also determining the inclination of the vector that relates the bright and dark contrast of the strain field (c). Note that the reduced-ROI around the defect used here measures 70 × 70 pixels.

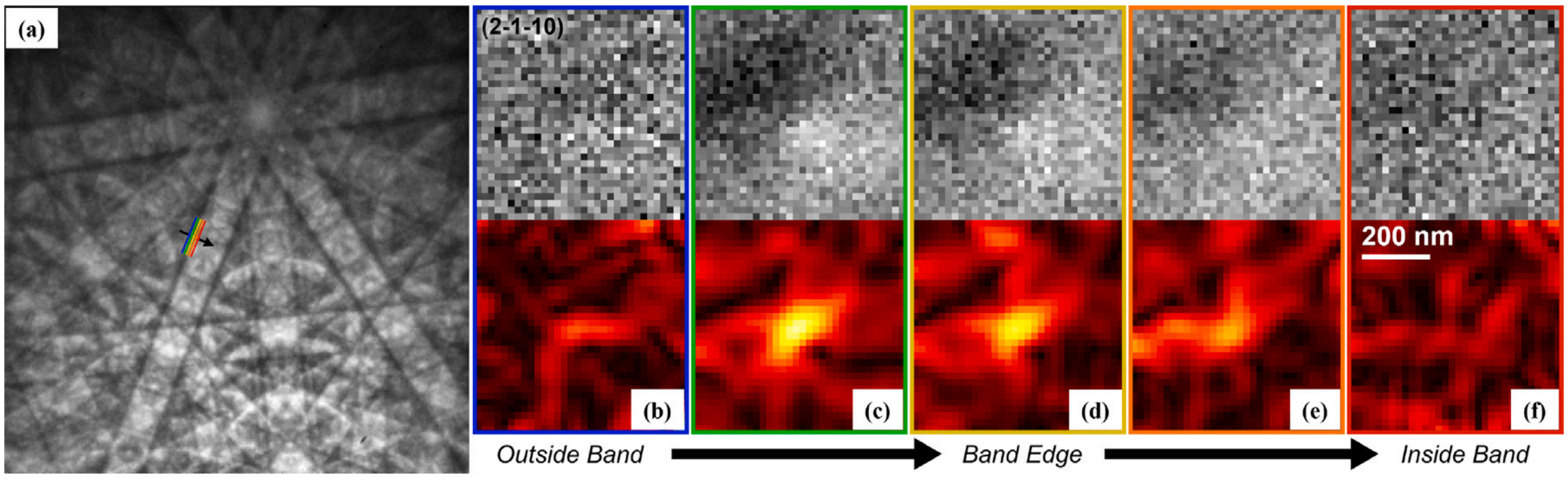

**Fig. 8.** Results of the convolutional kernel approach applied to the virtual images derived from the five OAVAs placed across the Kikuchi band edge indicated on the EBSD pattern in (a). (b–f) Top: virtual images corresponding to different OAVA positions; Bottom: results of the convolutional kernel approach. (d) Virtual image generated from the OAVA positioned precisely at the band edge. The reduced-ROI around the defect in this case measures 30 × 30 pixels. Note that the color bar in (b-f) is the same as in Fig. 7b.

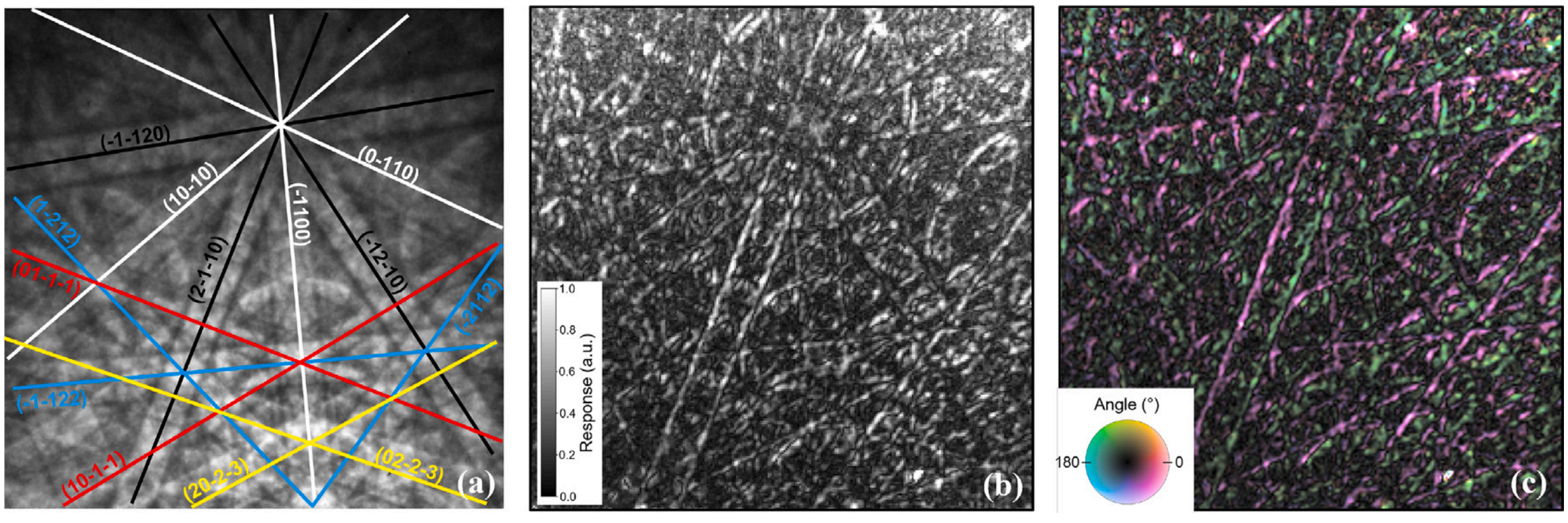

**Fig. 9.** The convolutional kernel approach applied to the full set of virtual images in Fig. A.1, derived from all 254 × 254 OAVA positions that correspond to the distinct crystallographic directions in the EBSD pattern. (a) Indexed EBSD pattern highlighting the principal Kikuchi bands. (b) Peak contrast gradient (PCG) map, where each pixel represents the maximum defect contrast intensity extracted from the virtual image generated at the corresponding EBSP location (and hence diffraction condition). (c) Composite map obtained by overlaying the PCG map with the peak contrast gradient direction (PCGD) map.

| | X | X | | X | X | X | X | X | X | X |
|---|---|---|---|---|---|---|---|---|---|---|
| **g/b** | **a/3[11-20]** | **a/3[1-210]** | **a/3[-2110]** | **[0001]** | **a/3[11-23]** | **a/3[1-213]** | **a/3[-2113]** | **a/3[-1-123]** | **a/3[-12-13]** | **a/3[2-1-13]** |
| [-1-120] | -2 | 1 | 1 | 0 | -2 | 1 | 1 | 2 | -1 | -1 |
| [2-1-10] | 1 | 1 | -2 | 0 | 1 | 1 | -2 | -1 | -1 | 2 |
| [-12-10] | 1 | -2 | 1 | 0 | 1 | -2 | 1 | -1 | 2 | -1 |
| [10-10] | 1 | 0 | -1 | 0 | 1 | 0 | -1 | -1 | 0 | 1 |
| [-1100] | 0 | -1 | 1 | 0 | 0 | -1 | 1 | 0 | 1 | -1 |
| [0-110] | -1 | 1 | 0 | 0 | -1 | 1 | 0 | 1 | -1 | 0 |
| [01-1-1] | 1 | -1 | 0 | -1 | 0 | -2 | -1 | -2 | 0 | -1 |
| [10-1-1] | 1 | 0 | -1 | 1 | 0 | -1 | -2 | -2 | -1 | 0 |
| [1-212] | -1 | 2 | -1 | 2 | 1 | 4 | 1 | 3 | 0 | 3 |
| [-1-122] | -2 | 1 | 1 | 2 | 0 | 3 | 3 | 4 | 1 | 1 |
| [-2112] | -1 | -1 | 2 | 2 | 1 | 1 | 4 | 3 | 3 | 0 |
| [02-2-3] | 2 | -2 | 0 | -3 | -1 | -5 | -3 | -5 | -1 | -3 |
| [20-2-3] | 2 | 0 | -2 | -3 | -1 | -3 | -5 | -5 | -3 | -1 |

**Fig. 10.** Approach for Burgers vector determination for a defect in wurtzite GaN via OAVA approach. (Left) PCGD map showing the centerline of the bands used for analysis. (Right) Truth table listing the principal accessible **g** vectors and possible Burgers vector **b** families. Each cell in the table represents the scalar product **g** · **b**, indicating whether the dislocation would be visible (nonzero; green-colored cells) or invisible (zero; red-colored cells) under the corresponding diffraction condition. The arrows on the left highlight the indices of the Kikuchi bands in the EBSP, each marked with a corresponding color. The "x"s above the columns indicate Burgers vectors ruled out as potential candidates, consistent with the rationale presented in the main text. The visibility and invisibility conditions across specific band edges suggest the dislocation Burgers vector to be $a/3[\bar{2}110]$.

Fig. 10 overlays the centerline of the indexed bands considered for the determination of the Burgers vector, accompanied by a table (also known as a truth table) listing the principal **g** vectors accessible within the covered reciprocal space in the EBSP, as well as the three candidate families of Burgers vectors **b** for wurtzite GaN. Each cell in the table represents the scalar product **g** · **b**, indicating whether the dislocation would be visible (nonzero) or invisible (zero) under the corresponding diffraction condition. The defect invisibility along the $(0\bar{1}10)$ band edges substantially down-selects the possible Burgers vectors, as it is characteristic of only a few possible dislocation types (those not marked with a white "x" at the top of the columns in the truth table). This constraint narrows the possibilities to just four dislocation types. The pronounced visibility of the defect along the $(2\bar{1}\bar{1}0)$ band edges (and for many other bands) suggests that the defect is not a c-type dislocation, as indicated by the black "x" marked at the top of the corresponding column in the truth table. The visibility along the $(\bar{2}112)$ and the invisibility along the $(02\bar{2}\bar{3})$ band edges further narrow the possible dislocation types to a singular Burgers vector: $a/3[\bar{2}110]$. Notably, a further invisibility condition for this dislocation type is required along the $(01\bar{1}\bar{1})$ band edges, which appears to be satisfied (red band in Fig. 10), thus substantiating the previous determination.

To validate the results obtained using the OAVA approach, we quantified via TEM the prevalence of a-type dislocations in the wurtzite GaN sample in an electron-transparent TEM foil extracted from a representative region of the specimen, followed by a detailed dislocation characterization. Specifically, the bright-field (BF) STEM image presented in Fig. 11a, acquired along the $[11\bar{2}0]$ zone axis (Fig. 11b), reveals a high density of threading dislocations within the foil. By tilting the sample to a [0002] two-beam condition, a significant fraction of these dislocations vanish from contrast, as observed in the corresponding BF STEM image (Fig. 11c). In wurtzite GaN, dislocations that satisfy the invisibility criterion under this diffraction condition are exclusively a-type, denoted by the white arrows in Fig. 11a. Conversely, the remaining dislocations, highlighted with yellow arrows in Fig. 11c, correspond to either c-type or c+a-type dislocations. By quantifying the fraction of a-type dislocations relative to the total dislocation population, a density of ~64% is determined for a-type dislocations. Given that the vertical axis of the image aligns with the *c*-axis of the hexagonal lattice, it follows that the identified a-type dislocations exhibit an edge character, for which the **g** · **b** invisibility criterion holds rigorously. This result lends strong support to the Burgers vector determination achieved through the OAVA approach.

## 4. Further considerations and outlook

Dislocation-sensitive imaging has traditionally been the domain of ECCI and TEM, each offering deep insights into defect structures in crystalline materials. Yet, despite their proven capabilities, both methods remain inherently constrained in throughput and scalability. ECCI, while powerful for surface-sensitive dislocation imaging, requires painstaking sample alignment and tilting to achieve diffraction conditions favorable for defect contrast — an iterative and often unstable process, especially in polycrystalline materials. The necessity for tilt-induced beam alignment, the synchronization of EBSD maps with ECPs, and the absence of unified software infrastructures for stage coordination collectively hinder ECCI's adoption for large-area analyses. TEM, by contrast, delivers atomic-resolution insights and allows for robust Burgers vector determination, but is equally hindered by its demand for extensive sample preparation, limited field of view, and the need to acquire multiple images under distinct diffraction conditions to resolve defect character — rendering it impractical for high-throughput or statistically significant investigations.

Within the SEM domain, beyond ECCI, recent works utilizing high-resolution EBSD (HR-EBSD) and HR-TKD have also shown the ability to detect and characterize individual dislocations [35,36]. HR-EBSD, particularly with advanced noise reduction methods, has been validated against TEM and ECCI for resolving individual dislocations and identifying their Burgers vectors [35]. Complementarily, HR-TKD, while requiring electron-transparent thin foils akin to TEM specimens, has demonstrated precise mapping of the full lattice strain tensor and rotations around single dislocations, enabling direct Burgers vector determination [36]. Recent advances in the post-processing of EBSD data have also significantly enhanced the visualization of defects. Notably, Hiller et al. [37] introduced virtual diode (VD) imaging, in which the pixelated EBSD detector is segmented into smaller "virtual diodes" to capture local variations in backscattered electron intensity. Center of mass (COM) imaging was also employed to track shifts in the centroid of the intensity distribution, while a hybrid approach, termed VDCOM, was developed and shown to offer improved signal-to-noise ratios compared to the individual techniques. McAuliffe et al. [38] further proposed spherical-angular dark field imaging, a method that synthesizes virtual microstructural contrast by summing intensity within one Bragg angle of a projected crystallographic plane on the diffraction sphere, thereby leveraging specific diffraction conditions to reveal subtle crystallographic features. More recently, Calvat et al. [39] developed a data-driven approach that encodes entire Kikuchi patterns into low-dimensional latent spaces using machine learning, subsequently

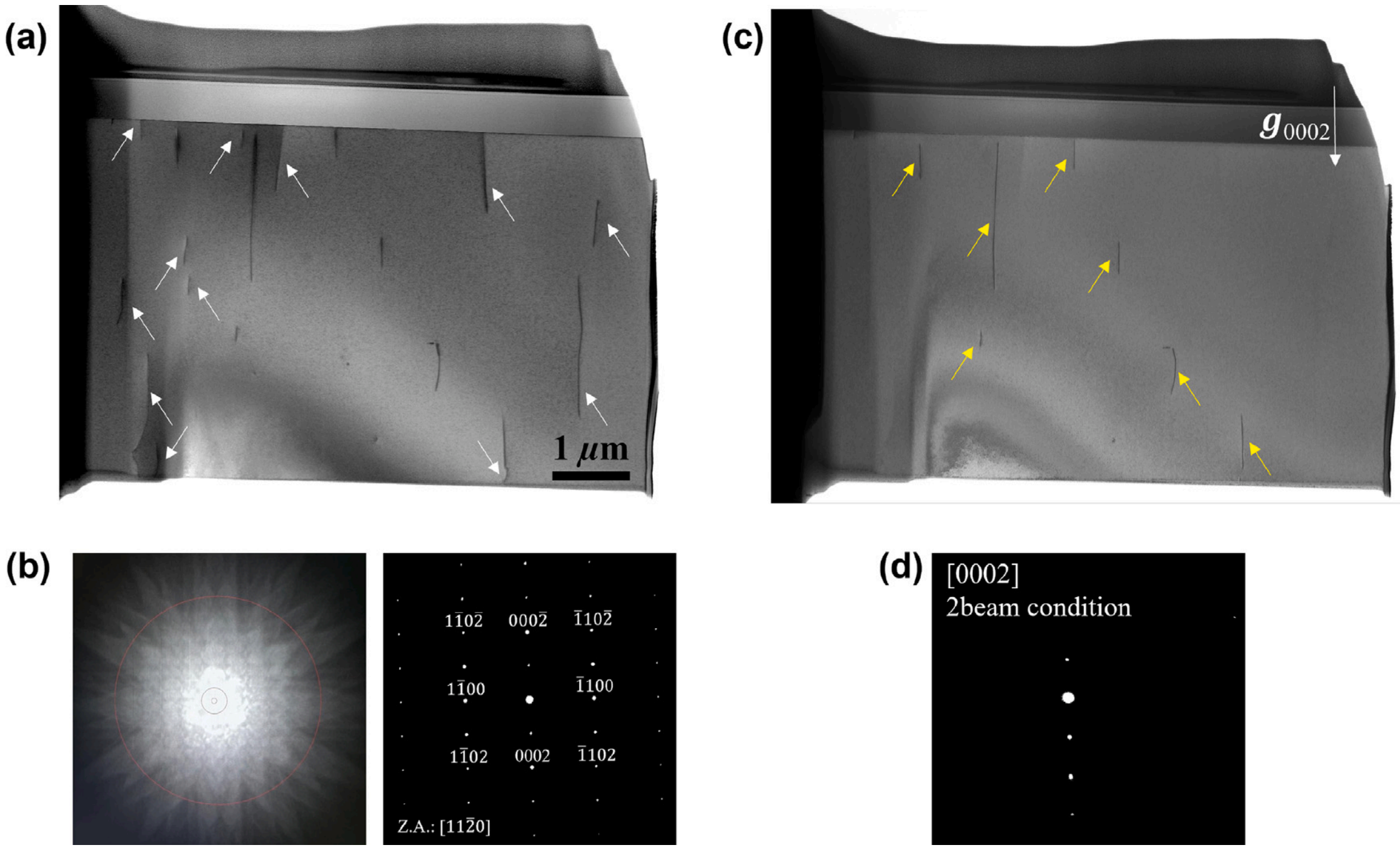

**Fig. 11.** TEM analysis of the electron-transparent GaN foil prepared via FIB. (a) Bright-field (BF) STEM image acquired along the [$11\bar{2}0$] zone axis, as referenced in (b), revealing all threading dislocations within the foil. (c) BF STEM image obtained under the [0002] two-beam condition, as referenced to (d), selectively highlighting non-a-type threading dislocations (yellow arrows). Dislocations that are no longer visible in (c) compared to (a) are of the a-type (white arrow in (a)).

mapping these latent features to capture and visualize microstructural heterogeneity.

In addition to these methods, the framework introduced here, based on orientation-adaptive virtual apertures integrated within a 4D-EBSD acquisition scheme, also redefines the possibilities for defect imaging. By leveraging the simultaneous collection of large areas of reciprocal space recorded in a single EBSD scan, OAVA enables the generation of virtual images tied to specific diffraction conditions without mechanical tilting and over large sample areas. This capability parallels the core principles of diffraction-contrast TEM imaging but eliminates the labor-intensive steps associated with sample preparation, sample alignment and multiple zone axis acquisition. The ability to dynamically align virtual apertures with the local crystallographic orientation unlocks contrast from strain fields surrounding dislocations, offering a novel route to defect imaging at scale.

Crucially, this study demonstrates that 4D-EBSD with OAVA not only provides qualitative visualization of dislocations, but also lays the foundation for dislocation characterization via visibility/invisibility criteria under different diffraction conditions. Sampling multiple **g** vectors simultaneously from a single scan offers a substantial improvement in efficiency over traditional two-beam methods used in TEM or even 4D-STEM approaches requiring sample tilting for full dislocation analysis. Moreover, the integration of direct electron detectors amplifies the fidelity of virtual imaging by enhancing low-voltage sensitivity, reducing interaction volumes, and improving spatial resolution — critical for resolving surface-penetrating dislocations and minimizing contrast loss in noisy patterns.

In the preceding Section 3.2.2, a potential strategy for identifying the Burgers vector of a selected defect was proposed, highlighting the capability of the OAVA approach for defect analysis. The results presented here are intended to be an initial estimate of the Burgers vector. Indeed, the determination of dislocation character — beyond presence and morphology — requires careful consideration. The contrast behavior in OAVA imaging is governed by a multifaceted set of parameters: the size and shape of the virtual aperture, camera geometry, sample tilt, detector resolution, background correction strategies, the accuracy of crystallographic indexing, excess-deficiency effects, interaction volume, surface relaxation phenomena associated with surface-penetrating dislocations, among others. For example, a narrow aperture offers high angular selectivity but is prone to noise; a broader aperture increases signal strength but sacrifices directionality, sampling neighboring crystallographic directions. Similarly, a shorter camera length increases the angular coverage per pixel, thereby reducing angular resolution, but increases the amount of reciprocal space sampled. These trade-offs necessitate optimization tailored to material symmetry, defect density, and pattern quality. In materials with low symmetry or pseudosymmetry, such as those highlighted in recent EBSD studies on monoclinic phases [18], minor orientation uncertainties can propagate into significant deviations in reconstructed contrast, underscoring the importance of precise calibration. Finally, approaches to improving signal-to-noise ratios in virtual imaging by summing multiple virtual apertures along a specific diffraction band edge need to be carefully balanced with the influence of crossing bands and other dynamical effects.

Taken as a whole, this work represents a foundational advance — a shift from diffraction-limited, tilt-dependent dislocation imaging towards a high-throughput, orientation-specific, and computationally adaptable framework for defect analysis. While a definitive validation of defect character via one-to-one correlation with TEM remains an avenue for future study, the demonstrated capability of 4D-EBSD and OAVA to reveal dislocation structures without the constraints of mechanical tilting or sequential imaging positions it as a powerful addition to current defect characterization methods.

## 5. Conclusions

This study introduces orientation-adaptive virtual apertures (OAVA) as a novel method for dislocation imaging within a 4D-EBSD approach. By dynamically aligning virtual apertures in reciprocal space with the local crystallographic orientation, OAVA enables post-acquisition generation of defect-sensitive contrast within virtual images, akin to diffraction-contrast TEM, but without the need for mechanical tilting or sample realignment. This capability facilitates the simultaneous visualization of dislocations across a wide range of diffraction conditions from a single EBSD scan, offering a scalable and high-throughput route to defect analysis. Key findings are listed below:

- By adapting virtual apertures to the local orientation of each grain, identical imaging conditions can be virtually applied across a polycrystalline field of view. This enables consistent contrast for dislocations in differently oriented grains, overcoming long-standing limitations in large-area defect analysis.
- Application of OAVA to an EBSD map of GaN revealed that dislocation visibility and invisibility vary systematically with diffraction conditions. The defect contrast intensity increases progressively from the center of a Kikuchi band towards its edge, reaching a maximum just outside the band edge, where *anomalous transmission* occurs (i.e., positive excitation error, $s_{\mathbf{g}} > 0$). Additionally, the contrast inverts across the width of the Kikuchi band. These findings are consistent with theoretical expectations, validating the method's capability for defect analysis without the need for sample tilting or sequential acquisitions.
- A convolution kernel method was implemented to automate the identification of dislocation contrast. This tool quantifies gradient intensities and directionality associated with the strain fields around threading dislocations, opening pathways to automated and scalable defect mapping.
- A truth table was used to relate specific contrast signatures — such as visibility/invisibility along different Kikuchi bands (i.e., diffraction conditions, **g** vectors) — to underlying dislocation characteristics. This logical framework serves as a basis for future rule-based or machine learning–driven approaches to classify dislocation types based on virtual imaging data.
- The accuracy and robustness of the OAVA approach are inherently tied to the quality of the EBSD patterns and the precision of both crystallographic orientation and pattern center determination and benefited from the of a modern direct electron detector in the current study. High-fidelity input data is critical to faithfully reconstruct diffraction contrast and to ensure reliable defect visibility assessments. As such, future applications of this technique will benefit significantly from continued advancements in detector performance, calibration strategies, and noise reduction algorithms.

As the first implementation of orientation-adaptive virtual apertures for defect imaging in EBSD, this work represents only the initial step in a broader development path. Much remains to be explored in terms of optimizing virtual aperture design, quantifying defect contrast, and extending the technique to more complex materials systems. Nonetheless, the framework introduced here represents a powerful addition to the current suite of defect characterization techniques, bridging the gap between high-resolution but low-throughput methods and scalable, non-destructive alternatives. By enabling post-acquisition imaging of dislocations under multiple diffraction conditions from a single EBSD map, OAVA opens new avenues for automated, high-throughput, and statistically meaningful defect characterization.

## CRediT authorship contribution statement

**Nicolò M. della Ventura:** Writing – original draft, Visualization, Software, Methodology, Investigation, Formal analysis, Data curation, Conceptualization. **James D. Lamb:** Writing – review & editing, Visualization, Software, Methodology, Investigation, Formal analysis, Data curation, Conceptualization. **William C. Lenthe:** Writing – review & editing, Visualization, Software, Methodology, Investigation, Formal analysis, Data curation, Conceptualization. **McLean P. Echlin:** Writing – original draft, Visualization, Software, Methodology, Investigation, Formal analysis, Data curation, Conceptualization. **Julia T. Pürstl:** Writing – review & editing, Visualization, Methodology, Investigation. **Emily S. Trageser:** Writing – review & editing, Investigation. **Alejandro M. Quevedo:** Writing – review & editing, Investigation. **Matthew R. Begley:** Writing – review & editing, Supervision, Resources, Project administration, Methodology, Funding acquisition, Conceptualization. **Tresa M. Pollock:** Writing – review & editing, Supervision, Resources, Project administration, Methodology, Funding acquisition, Conceptualization. **Daniel S. Gianola:** Writing – review & editing, Supervision, Resources, Project administration, Methodology, Funding acquisition, Conceptualization. **Marc De Graef:** Writing – original draft, Visualization, Supervision, Software, Resources, Project administration, Methodology, Investigation, Funding acquisition, Formal analysis, Data curation, Conceptualization.

## Declaration of competing interest

The authors declare that they have no known competing financial interests or personal relationships that could have appeared to influence the work reported in this paper.

## Acknowledgments

This research was supported by funds from the UC National Laboratory Fees Research Program of the University of California, USA, Grant Number L22CR4520, funds from the Army Research Laboratory accomplished under Cooperative Agreement Number W911NF-22-2-0121, and funds from the Department of Energy - National Nuclear Security Administration, USA under Award Number DE-NA0004152. The views and conclusions contained in this document are those of the authors and should not be interpreted as representing the official policies, either expressed or implied, of the Army Research Laboratory or the U.S. Government. The U.S. Government is authorized to reproduce and distribute reprints for Government purposes notwithstanding any copyright notation herein. The authors also acknowledge the NSF MRI instrumentation, USA grant No. 2117843. The research reported here made use of the shared facilities of the Materials Research Science and Engineering Center (MRSEC) at UC Santa Barbara: NSF DMR–2308708. The UC Santa Barbara MRSEC is a member of the Materials Research Facilities Network (www.mrfn.org). MDG acknowledges financial support from a National Science Foundation, USA grant (DMR-2203378), use of the computational resources of the Materials Characterization Facility at Carnegie Mellon University, USA supported by grant MCF-677785, as well as support from the John and Claire Bertucci Distinguished Professorship in Engineering, USA. NMdV would like to acknowledge Bailey E. Rhodes and Leah H. Mills for their scientific inputs and general contributions during the course of this project.

## Appendix A

See Fig. A.1.

## Appendix B. Supplementary data

Supplementary material related to this article can be found online at https://doi.org/10.1016/j.ultramic.2025.114205.

## Data availability

Upon manuscript acceptance, the data will be published in compliance with FAIR principles at https://doi.org/10.34863/t39t-fj24

Dataset: Orientation-Adaptive Virtual Imaging of Defects using EBSD (Original data) (PARADIM Data Collective)

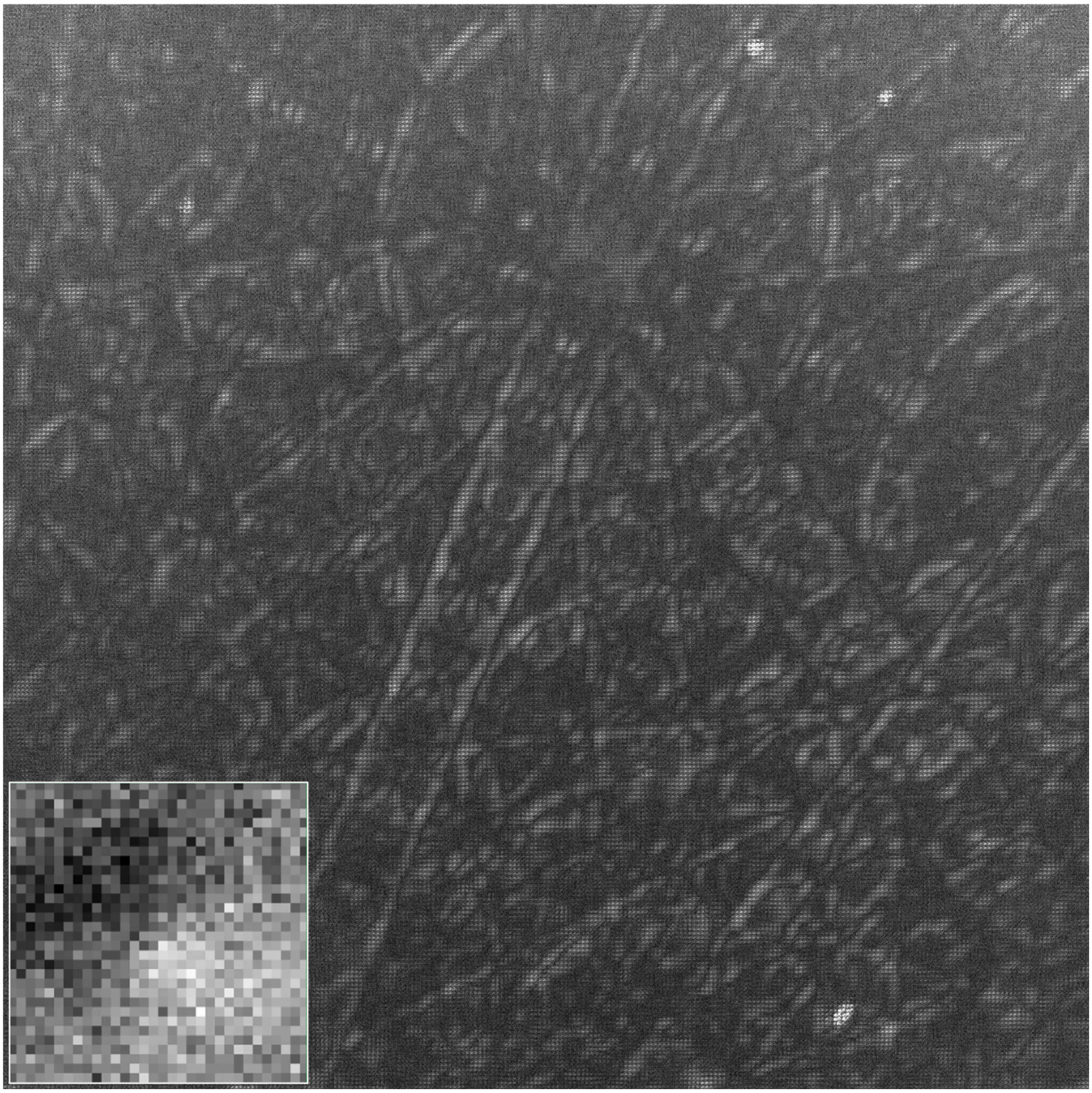

**Fig. A.1.** Mosaic of virtual images generated from the GaN datasets using a periodic array of 254 × 254 single-pixel-sized orientation-adaptive virtual apertures corresponding to all the accessible diffraction conditions from the EBSD pattern. One such virtual image constituting the mosaic is shown in the bottom-left corner of the figure. As the size of the ROI around the selected defect is 30 × 30 pixels, the overall size of the mosaic is 7620 × 7620 pixels. Virtual apertures positioned near the detector edges may produce undefined contrast in the resulting virtual images, as local misorientations (e.g., near defects) can displace portions of the aperture beyond the bounds of the sampled reciprocal space. These apertures are not shown in the current image but are included in the full-resolution, uncompressed image available for download in the supplementary materials ("Mosaic-256 × 256-OAVAs").